# The Impact of the Temporal Distribution of Communicating Civilizations on their Detectability

Amedeo Balbi[1]

**Abstract**

We use a statistical model to investigate the detectability (defined by the requirement that they are in causal contact with us) of communicating civilizations within a volume of the universe surrounding our location. If the civilizations are located in our Galaxy, the detectability requirement imposes a strict constraint on their epoch of appearance and their communicating lifespan. This, in turn, implies that the fraction of civilizations of which we can find any empirical evidence strongly depends on the specific features of their temporal distribution. Our approach shed light on aspects of the problem that can escape the standard treatment based on the Drake equation. Therefore, it might provide the appropriate framework for future studies dealing with the evolutionary aspects of the search for extraterrestrial intelligence (SETI). Key Words: Astrobiology—Extraterrestrial life—SETI—Complex life—Life detection—Intelligence

[1] Dipartimento di Fisica, Università degli Studi di Roma "Tor Vergata"
Via della Ricerca Scientifica, 00133, Roma, Italy
e-mail: balbi@roma2.infn.it

## 1. Introduction

In the past twenty years, the question "Are we alone?" has had increasingly better prospects of being scientifically investigated, due to the discovery of thousands of nearby exoplanetary systems, many of them of terrestrial size and in the habitable zone of their host star (Kane et al., 2016). In particular, the search for extraterrestrial intelligence (SETI) is going through a renaissance (for an up-to-date perspective, see Cabrol, 2016). Most discussions on extraterrestrial intelligence, building upon the pioneering work of Frank Drake and its famous equation (Drake, 1962), focus on estimating the average number of communicating species existing in a given volume of the universe surrounding our location. A well-known difficulty of this approach is that the key factors quantifying the probability that life might appear on a suitable planetary habitat and develop intelligence, communication and technology are largely uncertain and unpredictable. Furthermore, it has been noted (Ćirković, 2004) how a major shortcoming of most analyses based on the Drake equation is that they neglect the important role played by evolutionary processes, both astrophysical and biological.

In this paper, we propose a different approach, based on the application of Monte Carlo statistical techniques, to the problem of estimating the abundance of communicating civilizations that might be detected by SETI programs. While the use of numerical simulations within the SETI community has some past examples (e.g., Forgan & Rice 2010; Ćirković & Vukotić 2008), our method differs in making minimal assumptions, requiring just the fulfillment of a basic causal constraint. As such, our framework is flexible enough to be applicable to various scenarios: in particular, it is especially well-suited to treat the evolutionary aspects of SETI and exploring the impact of spatiotemporal dependencies on the final result.

**2. Method**

Our method consists of two parts. First, we notice that any civilization that can be in causal contact with us has to satisfy a condition involving its distance, its time of appearance and its communication lifespan; then, we simulate how the statistical distribution of these timescales impacts on the number of civilizations passing the criterion.

*2.1. Causal Requirement*

It is straightforward to write down the causal requirement that has to be met for an electromagnetic signal emitted by a communicating civilization to be detected by us. Let us call $t_c \geq 0$ the time in the past, measured from the present epoch (taken to be $t = 0$), when a given technological species started communicating (either intentionally or not) over interstellar distances. Let us also suppose that the same civilization has been broadcasting uninterruptedly over a lifespan $\tau \geq 0$, so that its communications ceased at time $t_{LS} = t_c - \tau$ (negative values mean that the civilization is still communicating). If the civilization is located at a distance $D$ from us, then we can receive its communication now, or in the future, only if its last signal was emitted after $D/c$, that is:

$$t_C - \tau \leq D/c \qquad (1)$$

(otherwise, the last signal reached us in the past, and we will never know).

We note, incidentally, that the above causal requirement for detectability is general enough to encompass any kind of information we can receive through the electromagnetic channel: this is not limited to radio messages (intentional or not) but includes, for example, the electromagnetic signatures of artifacts (see, e.g., Freitas & Valdes, 1985), astroengineering mega-structures (Dyson, 1960; see, also, Bradbury, Ćirković & Dvorsky, 2011), etc. Furthermore, it might easily be adapted to treat signals traveling at subluminal speed (see, e.g., Rose & Wright, 2004).

Despite its simplicity, inequality (1) has interesting implications. Notably, although no dependence is to be expected, a priori, between the communicating lifespan of an intelligent species and the time when it made the transition to radio technology, the two timescales cannot assume arbitrary values once we require that we can receive its signals. This introduces a selection effect on the kind of civilizations we might come in contact.

In particular, for any communicating civilization located within our Galaxy, the implied fine-tuning between the two timescales is rather severe. This can be seen as a direct consequence of the fact that the light crossing time of our Galaxy ($\sim 10^5$ years) is much smaller than its age ($T_G \sim 10^{10}$ years), so that the term on the right-hand side of (1) is very small compared to the typical age of stars. This will select for contact only those civilizations with $t_C$ and $\tau$ matching within a few orders of magnitude. Intuitively, this means that if a civilization appeared much earlier than ours, its communicating lifespan should also be much larger than ours if we have to receive its signals; conversely, shorter lifespans are allowed only if the civilization started communicating nearly simultaneously to ours. This already tells us something on the kind of civilization we might expect to find evidence of, no matter what their abundance is.

More generally, inequality (1) suggests that, given a total number of intelligent communicating species spread over the history of the Galaxy, the size of the sample that can be in causal contact with us will strongly depend on the properties of the distribution of their $t_C$ and $\tau$.

## 2.2. Statistical Modeling

In order to get some quantitative insight on this issue, we performed Monte Carlo simulations of $N_{tot}$ communicating civilizations characterized by random values of the variables $D$, $t_C$ and $\tau$, each drawn from its own probability distribution. For simplicity, we assumed the locations of civilizations to be uniformly distributed within a sphere of radius $R_E$ centered on our position, and extracted their random distances $D$ accordingly ($D$, of course, is not uniformly distributed). The choice of a probability distribution for $t_C$ and $\tau$ should, ideally, rely on the adoption of some predictive model (deriving from our understanding of astrophysical, biological or sociological processes). While some possibilities can be more motivated than others based on current knowledge (see Section 4 for a discussion of this issue), we decided here not to commit to a specific model, and rather tried different options, with the only purpose of illustrating their influence on the final result. The appearance times $t_C$ were drawn either from a uniform distribution in the range $(0, T_G)$, with $T_G = 10^{10}$ years, or from a truncated normal distribution in the range $(0, T_G)$, with given mean $\langle t_C \rangle$ and standard deviation $\sigma$. The values of $\langle t_C \rangle$ and $\sigma$ for the normal distributions were chosen to exemplify a case in which our civilization appearance time is "typical" ($\langle t_C \rangle$ = 1 Gy, $\sigma$ = 2 Gy) and two cases in which we appeared much later than most other civilizations ($\langle t_C \rangle$ = 4 Gy, $\sigma$ = 1 Gy and $\sigma$ = 2 Gy). The probability distribution of lifespans was modeled as an exponential distribution with given mean $\langle \tau \rangle$.

For each choice of a probability distribution for $t_C$ and $\tau$, we simulated the same total number of civilizations, $N_{tot} = 10^7$, then computed the number $N_D$ of civilizations satisfying condition (1) and evaluated a "detectable fraction", defined as:

$$f_D = N_D/N_{tot}$$

We repeated each simulation 100 times and averaged the resulting $f_D$ in order to get a more reliable estimate of its value.

## 3. Results

### *3.1. Detectable Fraction*

As a realistic case study, we chose to evaluate the expected $f_D$ within a spherical volume of radius $R_E = 10^3$ ly centered on our location. This is a scientifically well-motivated choice for various reasons. First, $10^3$ ly is roughly the depth that can be probed by present-day SETI surveys (see, e.g., Tarter, 2007), assuming that the strength of the signal is comparable to what our own civilization is capable of producing. Second, we now have increasingly accurate empirical constraints on the abundance and type of exoplanets just within the same volume (Batalha, 2014), resulting in reasonable statistical estimates of the fraction of planetary systems meeting the basic requirements for life (Petigura, Howard & Marcy, 2013) as well as of the potential targets for SETI searches (Harp et al, 2016). Third, there are strong theoretical arguments (Gonzales, Brownlee & Ward, 2001) for assuming that there is a region within the Milky Way more suitable for the appearance of complex life: astrophysical models (Lineweaver, Fenner & Gibson, 2004) estimate that this region, dubbed Galactic Habitable Zone (GHZ), encompasses an annulus of width ~2 kpc centered at a distance of ~8 kpc from the Galactic Center. Thus, the volume chosen for our simulations would fall well within the borders of the GHZ.

Figure 1 shows our results for $f_D$ as a function of the average communicating lifespan $\tau$, for the various choices described in the previous section for the statistical distribution of $t_c$. As expected, the fraction of detectable civilizations varies considerably. In particular, for the same average communicating lifespan, the temporal distribution for the appearance of civilizations has a strong effect on the final result.

*3.2. Comparison with the Drake Equation*

It is interesting to draw a connection between our approach and the standard one based on the Drake equation. First of all, we point out that the Drake equation can be seen simply as an application of a well-known result of queueing theory, Little's law (Little, 1961):

$$N = \lambda L$$

where $N$ is the long-term average number of items in a stable system, $\lambda$ is the long-term average item arrival rate and $L$ is the average time spent by each item in the system. In the case of the Drake equation, the system is a volume of space and the items are the communicating civilizations in it; $N$ is their average number, $\lambda$ is their average rate of appearance (usually written as the product of the expected frequencies of various astrophysical, biological and sociological occurrences), and $L$ is their average longevity. Note that the Little's law holds under the assumption of stationarity of the underlying stochastic process, which clearly cannot be justified, either empirically or from general principles, in the case under examination. In fact, evolutionary processes are *exactly* what we are interested in, but the structure of the formula, and the assumption it rests on, prevents one to incorporate them. Note, also, that the average operation has two different meanings at the two sides of the equation: on the left-hand side, it is an average over time; on the right-hand side, it is an average over the total number of items in the system (or a representative sample).

Now, $\lambda$ and $L$ are somehow related to the characteristic timescales of our model: $L$ is simply $\langle\tau\rangle$, but the connection between $\lambda$ and $t_C$ is not straightforward. In fact, the average involved in the estimate of $\lambda$ masks out any difference in the distribution of $t_C$: if $N_{tot}$ is the total number of civilizations ever appeared in the volume, then $\lambda = N_{tot}/T_G$, regardless of how the time of appearance is spread over $T_G$. Finally, since $N$ is an average over time, it is in general different from $N_D$: in particular, $N$ does not account for the travel time of electromagnetic signals. This is a negligible effect for small distances, but it can be relevant as one explores larger volumes of space. Putting this all together, we see that:

$$N_D = N f_D \frac{T_G}{\langle\tau\rangle}$$

For $D/c \ll T_G$, we expect that $f_D \approx \langle\tau\rangle/T_G$ if $t_C$ is uniformly distributed in the interval $(0, T_G)$. In this case, $N_D$ essentially converges to the estimate of the Drake equation, as one would expect. This is in agreement with our numerical results (see Figure 1). For different temporal distributions, however, the actual number of detectable civilizations could be substantially different from what estimated applying the Drake equation.

## 4. Discussion and Conclusions

Our results highlight the necessity, already noted elsewhere (Ćirković, 2004), of modeling the effect of evolutionary processes when attempting to estimate the number of communicating civilizations that might be in causal contact with us. While the Drake equation remains a very useful guide in identifying and analyzing the various factors involved in the problem and their relative probabilities, our approach could provide a better framework to perform a statistical analysis when taking into account evolutionary processes. In particular, we have shown that the Drake equation can misestimate the number of detectable communicating species if spatial and temporal dependencies are important. This should suggest some caution in the uncritical application of such equation. In this paper, we gave an exemplificative illustration of the effect of different probability distributions for the typical timescales involved in the problem. Ideally, one would want to arrive at a mathematical description of such distributions based on a modeling of the underlying astrophysical and biological processes. This points out to a possible future direction of investigation, in which an integrated evolutionary (and co-evolutionary) approach to the problem occupies a center stage (for more on this perspective, see, e.g., Chaisson, 2001; Maccone, 2011; Ćirković, 2012). This is a daunting task, and certainly beyond the scope of this paper. However, we can make some general considerations.

The average communicating lifespan has long been known to play a crucial role in the estimation of the expected number of existing civilizations at any given time. Thus, it is generally to be expected that $\langle \tau \rangle$ must exceed some minimal value in order to have at least one other intelligent species in the volume under consideration (for a recent discussion, see Wandel, 2015). Our results however suggest that the minimal lifespan will vary depending on the exact distribution of the time of appearance of communicating civilizations. Just to fix the ideas, let us suppose that, over the history of our Galaxy, there has been a total of $\sim 10^6$ communicating civilizations in a radius of $R_E = 10^3$ ly around us. We then need $f_D \gtrsim 10^{-6}$ to have at least one detectable civilization in the volume. Depending on the distribution of $t_C$ assumed in our simulations, this means an average lifespan as low as $10^3$ years or as high as $10^8$ years (the Drake equation value would be $\sim 10^4$ years): clearly, one expects that such lifespans have very different chances of being reached, with a consequently different impact on the prospect of success of SETI. Also, we point out that the end of the communicating period of a civilization does not necessarily coincide with the time of its biological extinction: it can occur earlier (if the civilization goes radio-quiet at some point of its technological history) or later (for example, through the construction of automated interstellar beacons (see, e.g., Benford, Benford & Benford, 2010) or the persistence of detectable astroengineering structures). It is also not inconceivable that in the same planetary locations multiple technological species may arise and disappear at different epochs, separated by "dark" periods. Anyway, currently, there is no way to make evidence-based predictions on plausible values of $\tau$.

The distribution of $t_C$ may have some better prospects of being modeled, based on astrophysical considerations coupled with reasonable guesses of biological timescales. Non-uniform temporal distributions for the appearance of complex life have been argued for: according to some estimates (Lineweaver, 2001; Lineweaver, Fenner & Gibson, 2004), 75% of planets that might have developed complex life would be orbiting stars much older than ours, with a peak at an age $\sim 10^9$ years older than the Sun (see, also, Behroozi & Peeples, 2015). Note that such temporal distribution would approximately correspond to the case $\langle t_C \rangle = 1$ Gy, $\sigma = 2$ Gy of one of our simulations. Other non-uniform probabilities have been explored in the literature, including the existence of some astrophysical mechanism which synchronizes the appearance of complex life in the Galaxy at late epochs (as in Annis, 1999), or a monotonically increasing likelihood for the appearance of life over cosmic time (Loeb, Batista & Sloan, 2016; Dayal, Ward & Cockell, 2016).

As a final note, we point out that the detectable fraction is also affected by the spatial distribution of communicating civilizations. Although we have not explicitly addressed the issue in this paper, this has at least two interesting aspects that might be worth investigating in future works. The first is relaxing the assumption of spatial uniformity of communicating civilizations, for example by exploring specific models of the GHZ. The second is to vary the extent of the radius under investigation. This impacts on the detectable fraction, because the travel time of electromagnetic signals grows with the distance of civilizations. In particular, as $D/c$ approaches $T_G$, the constraint on $t_C - \tau$ relaxes considerably: intuitively, we could pick up communications of long extinct civilizations ($t_{LS} \gg 1$) if they are very distant from us, even if their communicating lifespan is short. In contrast, the Drake equation is only estimating the number of civilizations which are communicating at present time, in general a much smaller number. As an illustration, if we extend our simulation to extra-galactic scales, $R_E = 10^9$ ly, and assume a uniform distribution in $t_C$, we obtain $f_D = 0.1$ for an average lifespan $\langle \tau \rangle = 10^3$ years: all else being equal, this implies a number of detectable civilizations a factor $10^6$ larger than the estimate of the Drake equation. Barring considerations on the strength of the signal required for detection, this might suggest to expand the search for signs of communicating civilizations over extra-galactic distances (see, e.g., Lubin, 2016; Lingam & Loeb, 2017; Ćirković, 2012) in order to significantly increase the chances of success.

**Author Disclosure Statement**

No competing financial interests exist.

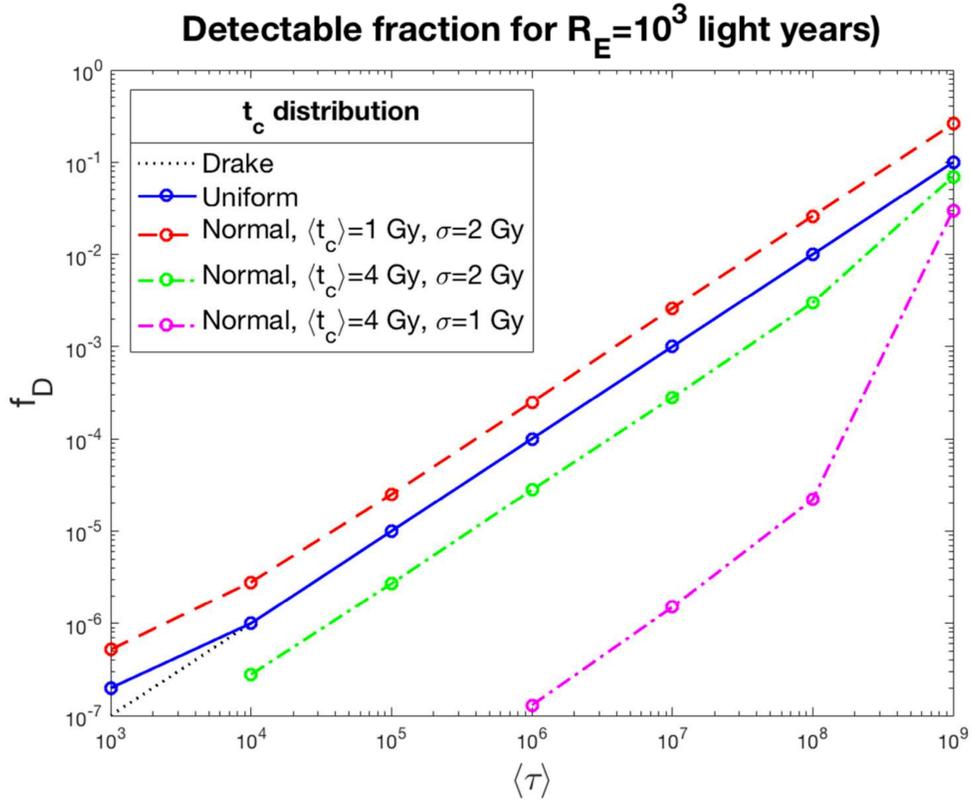

FIG.1. The fraction of detectable communicating civilizations in a volume of radius 1000 light years around our location, as a function of their average communication lifespan. Each line corresponds to a different probability distributions for the appearance time of civilizations, assuming that the same total number of civilizations appeared at random locations in the volume over the past 10 billion years. The dotted line represents the fraction of presently communicating civilizations, estimated by the Drake equation.